\documentclass[journal,10pt]{IEEEtran}
\ifCLASSINFOpdf
  \usepackage[pdftex]{graphicx}
  \usepackage{amsthm,amsmath,amssymb}
\usepackage{makecell}
\usepackage{mathrsfs}
  \usepackage{float}
  \usepackage{algorithm}
\usepackage{algpseudocode}
\usepackage{amsmath}
\usepackage{bm}
\usepackage{color}
\usepackage{comment}
\usepackage{booktabs}
 
\else
\fi
\usepackage{leftidx}
\usepackage{epstopdf}
\usepackage{amsmath}
\usepackage{subfigure}
\usepackage{makecell}
\usepackage{multirow}
\usepackage{soul, color, xcolor}
\usepackage{stfloats}
\usepackage{multirow}
\usepackage{cite}
\usepackage{enumitem}

\begin{document}

\title{WiCo-MG: Wireless Channel Foundation Model for Multipath Generation via Synesthesia of Machines
}

\author{
Zengrui Han, Lu Bai$^{*}$, Xuesong Cai, and Xiang Cheng$^{*}$

\thanks{Z. Han, X. Cai, and X. Cheng are with the State Key Laboratory of Photonics and Communications, School of Electronics, Peking University, Beijing, 100871, P. R. China (email: zengruihan@stu.pku.edu.cn, xuesong.cai@pku.edu.cn, xiangcheng@pku.edu.cn).}
\thanks{L. Bai is with the Joint SDU-NTU Centre for Artificial Intelligence Research (C-FAIR), Shandong University, Jinan, 250101, P. R. China (e-mail: lubai@sdu.edu.cn).}}


\maketitle

		\maketitle

\begin{abstract}
Precise modeling of channel multipath is essential for understanding wireless propagation environments and optimizing communication systems. 
In particular, sixth-generation (6G) artificial intelligence (AI)-native communication systems demand massive and high-quality multipath channel data to enable intelligent model training and performance optimization.
In this paper, we propose a wireless channel foundation model (WiCo) for multipath generation (WiCo-MG) via Synesthesia of Machines (SoM). 
To provide a solid training foundation for WiCo-MG, a new synthetic intelligent sensing-communication dataset for uncrewed aerial vehicle (UAV)-to-ground (U2G) communications is constructed.
To overcome the challenges of cross-modal alignment and mapping, a two-stage training framework is proposed.
In the first stage, sensing images are embedded into discrete-continuous SoM feature spaces, and multipath maps are embedded into a sensing-initialized discrete SoM space to align the representations. 
In the second stage, a mixture of shared and routed experts (S-R MoE) Transformer with frequency-aware expert routing learns the mapping from sensing to channel SoM feature spaces, enabling decoupled and adaptive multipath generation.
Experimental results demonstrate that WiCo-MG achieves state-of-the-art in-distribution generation performance and superior out-of-distribution generalization, reducing NMSE by more than 2.59~dB over baselines, while exhibiting strong scalability in model and dataset growth and extensibility to new multipath parameters and tasks.
Owing to higher accuracy, stronger generalization, and better scalability, WiCo-MG is expected to enable massive and high-fidelity channel data generation for the development of 6G AI-native communication systems.

\end{abstract}

\begin{IEEEkeywords}
Wireless channel foundation model (WiCo), Synesthesia of Machines (SoM), multipath generation (MG), SoM feature spaces, mixture of shared and routed experts (S-R MoE).
\end{IEEEkeywords}
\IEEEpeerreviewmaketitle

\section{Introduction}

\IEEEPARstart{F}{or}
wireless communication systems, both a thorough understanding of channel characteristics and accurate channel modeling are fundamental to efficient system design and algorithm development~\cite{cheng2022channel}. 
Among diverse channel characteristics, channel small-scale fading caused by multipath propagation is of particular importance, since it embodies the intricate interactions between electromagnetic waves and the surrounding environment~\cite{scatterering}.
These interactions lead to rapid temporal and spatial fluctuations in the received signal, which significantly affect instantaneous performance indicators, such as signal-to-noise ratio (SNR) and bit error rate (BER).
Therefore, a precise modeling of the multipath propagation is indispensable for the development of advanced transmission strategies and the assurance of robust system performance in dynamic and heterogeneous environments~\cite{cai4}.


From first-generation (1G) to fifth-generation (5G), the research on channel multipath fading has been extensively pursued, and substantial channel models have been proposed to provide an indispensable validation platform for system design and algorithm development.
In general, two main approaches, i.e., deterministic and stochastic channel modeling, are adopted.
Stochastic channel models~\cite{yang2019cluster,cai3} generate channel parameters in a stochastic manner with moderate accuracy and relatively low computational complexity.
{\color{black}Considering this trade-off between accuracy and complexity, standardized models, such as 3GPP TR 38.901 channel model~\cite{generation2019technical} and QuaDRiGa channel model~\cite{QuaDRiGa-zw}, also adopt stochastic channel modeling approaches.}
Deterministic channel models~\cite{qd1,qd2}, such as ray-tracing (RT), explicitly replicate the physical propagation process in a detailed environment with considerable computational complexity.
Overall, these approaches have sufficiently supported conventional system design from 1G to 5G.
Future sixth-generation (6G) systems are expected to embed artificial intelligence (AI) into the core algorithm development, giving rise to AI-native 6G systems.
In this paradigm, system performance is fundamentally bounded by the availability of massive and high-fidelity training data~\cite{tnse,LMMs}.
To support this demand, abundant channel data is required that can accurately capture multipath fading characteristics, including the continuous evolution, birth-death process, and non-stationarity, together with the critical features in the delay and angular domains.
However, due to the excessive complexity and idealized assumptions of deterministic models, and the limited precision of stochastic methods, these approaches remain inadequate for generating the massive and high-quality channel data required by AI-native 6G systems.

To address the limitations of conventional modeling approaches, the generative capabilities of advanced AI techniques offer a promising alternative for massive and high-fidelity channel data generation~\cite{AImodeling}.
A conceptually intuitive solution is to augment existing high-precision radio frequency (RF) channel data using generative AI techniques, including machine learning and deep learning.
The authors in~\cite{cluster3} generated trajectory-consistent electromagnetic clusters using a Kuhn-Munkres assignment algorithm based on real-world vehicle-to-vehicle (V2V) channel data.
{\color{black}Leveraging Wasserstein generative adversarial networks (GAN) with gradient penalty (WGAN-GP), the developed model in~\cite{gene1} learned from measured V2V channel data and generates an identical statistical distribution.}
To further consider uncrewed aerial vehicle (UAV) communications, the authors in~\cite{gene2} proposed a generative neural network approach that leverages a two-stage architecture trained on RT data to generate millimeter-wave (mmWave) UAV channel parameters.
However, the aforementioned methods remain constrained by two critical limitations. 
First, exclusive reliance on channel measurements~\cite{cluster3,gene1} requires stringent delay and angular resolution, placing heavy demands on system bandwidth and array size~\cite{localization,cai2}. 
Second, the lack of explicit physical environment representation~\cite{cluster3,gene1,gene2} reduces interpretability.
Moreover, it limits the generalization capability of these models to diverse or dynamic propagation scenarios.

With the advancement of 6G, intelligent agents are expected to possess comprehensive environmental awareness through multi-modal sensing technologies such as red-green-blue (RGB) cameras and light detection and ranging (LiDAR). 
Inspired by synesthesia of human, Synesthesia of Machines (SoM)~\cite{som} was proposed to enable the intelligent integration of multi-modal sensing and communication.
This perspective has opened up new avenues, i.e., cross-modal generation, where rich and easily accessible sensing data serves as a basis for synthesizing wireless channel data.
{\color{black}For instance,~\cite{vision} employed sensing data collected by cameras to generate point-to-point path loss utilizing a convolutional neural network (CNN)-based model, while~\cite{visionlidar} generated path loss distributions from RGB images, point clouds, and GPS information.
Moreover,~\cite{smallscale} utilized LiDAR and panoramic camera to generate root-mean-square (RMS) delay spread, angular spread, and Rician \textit{K} factor.}
However, the above studies~\cite{vision,visionlidar,smallscale} are limited to large-scale fading or small-scale statistical characteristics, rather than detailed geometric multipath parameters, which fundamentally reflects the essence of wireless propagation.
Meanwhile, these methods rely on lightweight deep learning models, which limit their capacity to generalize across diverse and dynamic scenarios expected in 6G AI-native communication systems.
As large language models (LLMs) enable the data generation in various fields~\cite{image,video,table,time} and exhibit significant adaptability on cross-modal tasks~\cite{LLM4CP,CSILLM,beam,feedback,estimation} through fine-tuning, recent studies have explored the potential of LLMs in cross-modal channel data generation.
In~\cite{LLM4SG}, our previous work adapted LLM to generate electromagnetic scatterers from LiDAR point clouds and demonstrated improvements in generation accuracy and generalization capability.
However, directly relying on LLMs remains limited in several respects.
On one hand, the pretraining priors of LLMs are not inherently aligned with the principles of electromagnetic propagation, which compromises the physical plausibility and fidelity of the generated channel data. 
Moreover, LLMs exhibit limited controllability in modeling communication-specific conditions such as operating frequency bands and environmental scenarios. 
On the other hand, the substantial statistical disparity between sensing data and fine-grained multipath parameters poses significant challenges for end-to-end fine-tuning, resulting in unstable cross-modal alignment.
{\color{black}Therefore, beyond leveraging the cross-modal representation capabilities of large models, it is essential to develop a wireless channel foundation model (WiCo) that integrates specialized architectural designs and domain knowledge grounded in sensing and wireless communications. 
Meanwhile, WiCo is trained from scratch on large-scale datasets encompassing both sensing and communication modalities, enabling it to learn the intrinsic mapping from sensing to multipath parameters.
WiCo exhibits higher accuracy, stronger generalization, and better scalability, thereby enabling massive and high-fidelity channel data generation for the development of 6G AI-native communication systems.}

\begin{figure}[t]
\centering
\includegraphics[width=3.5in]{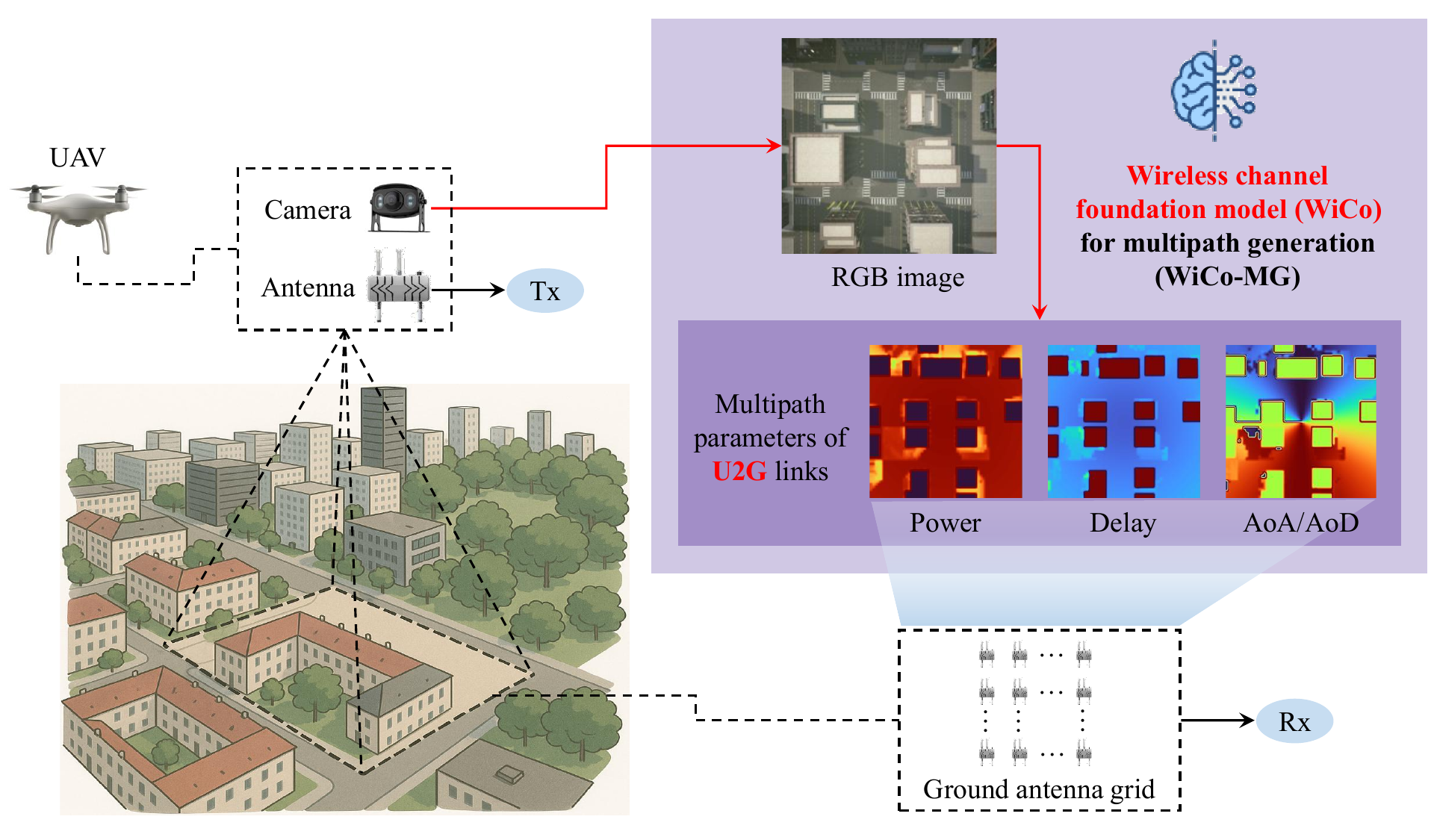}
\caption{Conceptual illustration of the proposed WiCo-MG, which generates multipath parameters of U2G links from RGB images obtained from UAV in urban scenarios.}
\label{story}
\end{figure}

To fill this gap, the first WiCo for multipath generation (WiCo-MG) from RGB images is proposed.
To be specific, RGB images and multipath data are meticulously embedded into sensing SoM feature space and channel SoM feature space, respectively, in a manner that preserves semantic and structural consistency.
Subsequently, a mixture of shared and routed experts (S-R MoE) Transformer is employed to establish the mapping between the sensing and channel SoM feature spaces, while simultaneously enabling the decoupled generation of distinct multipath channel parameters.
WiCo-MG is validated in a typical 6G scenario, namely UAV-to-ground (U2G) communications, as shown in Fig.~\ref{story}, whose complex and highly three-dimensional (3D) dynamic nature poses significant challenges for cross-modal data generation.
The major contributions and novelties of this paper are summarized as follows.

\begin{enumerate}
    \item To address the pressing demand for massive and generalizable data in AI-native 6G communication systems, we propose WiCo-MG, which enables cross-modal generation of multipath channel parameters directly from perceptual images for the first time, as shown in Fig.~\ref{story}. WiCo-MG demonstrates broad adaptability across diverse scene conditions, including scenario types, communication frequency bands, and UAV flight altitudes.
    \item A new synthetic intelligent sensing-communication dataset for U2G communications is constructed, which jointly captures sensing and geometrical channel multipath parameters. 
    {\color{black}The dataset spans multiple scenarios (e.g., crossroad and wide-lane), multiple frequency bands (e.g., 1.5 GHz, 5.9 GHz, 15 GHz, and 28 GHz), and multiple flight altitudes (e.g., 50 m, 70 m, 80 m, 200 m, 250 m, and 300 m) to meet various application scenario requirements and diverse communication conditions.}
    Moreover, the dataset includes 23.52\,k RGB images and 0.13 billion sets of multipath parameters, i.e., power, delay, angle of departure (AoD), and angle of arrival (AoA).
    \item A two-stage training framework is further developed to overcome key challenges in cross-modal representation alignment, multipath parameter disentanglement, and cross-frequency generalization. 
    In the first stage, RGB images are embedded into both discrete and continuous sensing SoM feature spaces via vision Transformer (ViT)-VQGAN and SigLIP-2, enabling the image tokens to support both generation and semantic understanding. 
    Simultaneously, multipath parameters are embedded into a discrete channel SoM feature space using a ViT-VQGAN, with its codebook initialized from the sensing side to explicitly reduce the representational gap between sensing and channel modalities. 
    In the second stage, a hybrid token-wise and task-wise S-R MoE Transformer is adopted to establish the mapping from sensing to channel SoM feature spaces, enabling the decoupled generation of multiple channel parameters. 
    Additionally, frequency embeddings are introduced into the MoE gating mechanism, allowing expert routing to adapt dynamically to communication frequencies.
    \item 
    {\color{black}Experimental results demonstrate that WiCo-MG achieves state-of-the-art in-distribution generation performance and superior out-of-distribution generalization, yielding an NMSE reduction of more than 2.59~dB over baselines including the fine-tuned LLM and conventional deep learning approaches. 
    Moreover, the model exhibits strong scalability with increasing model and dataset sizes, as well as significant extensibility to new multipath parameters and channel-related tasks.
    Furthermore, unlike standardized stochastic models such as 3GPP TR 38.901, WiCo-MG can generate scenario-specific and fine-grained multipath channels that better capture the physical characteristics of real environments.}
\end{enumerate}

The remainder of this paper is organized as follows.
Section II introduces the synthetic intelligent sensing-communication dataset in U2G communications.
The architecture and design of WiCo-MG are provided in Section III.
In Section IV, simulation settings are illustrated, and then the performance of the proposed WiCo-MG is evaluated.
Finally, Section V draws the conclusion.

\section{Dataset Construction}

In this section, a new synthetic intelligent sensing-communication dataset in U2G communications is constructed for multiple U2G scenarios with multiple frequency bands and multiple flight altitudes.
The constructed dataset contains RGB images and geometrical channel parameters, such as power, delay, and angle of multipath components.
To achieve the synchronous collection of the sensing and communication data, two high-fidelity simulation software, i.e., Wireless InSite~\cite{InSite} and AirSim~\cite{shah2018airsim}, are intelligently incorporated as an integrated platform~\cite{synthsom} to achieve in-depth integration and precise alignment.
Wireless InSite employs the RT technology to mimic radio wave propagation and collect channel data.
AirSim is an open-source platform developed on Unreal Engine and is used to collect high-fidelity sensing data, such as RGB images. 
The constructed dataset is designed to simulate diverse U2G communication conditions, which are challenging to capture in real-world measurements. 
It covers various scenarios, including crossroad and wide-lane, employs 1.5 GHz, 5.9 GHz, 15 GHz, and 28 GHz frequency bands, and accounts for different flight altitudes of UAVs.
Specific scenario configurations and data collection are given as follows.

\subsection{Scenario Configuration}

\begin{figure*}[t]
\centering
\includegraphics[width=\textwidth]{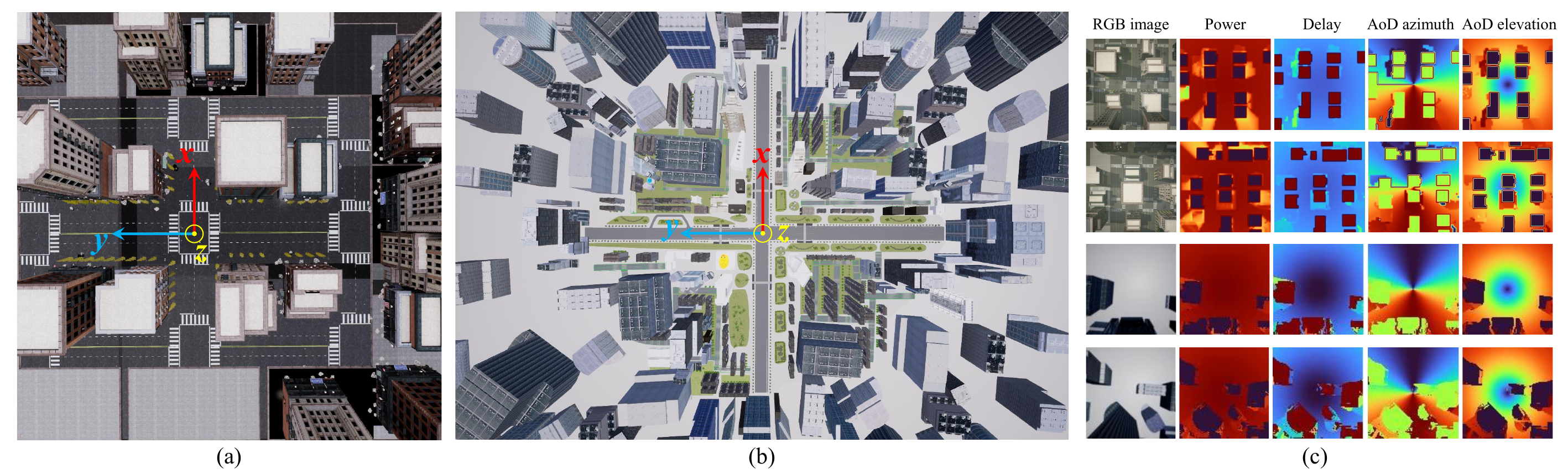}
\caption{Overview of the constructed dataset.
(a) Crossroad scenario. (b) Wide-lane scenario. (c) Dataset samples from the two scenarios.}
\label{dataset}
\end{figure*}

To collect multi-modal sensing and communication data across diverse scenarios and communication conditions, aligned initial environments are constructed in AirSim and Wireless InSite.
Two typical U2G scenarios, i.e., an urban crossroad and an urban wide-lane, are chosen, as shown in Fig.~\ref{dataset}.
The crossroad scenario, produced by PurePolygons, simulates bidirectional intersections, while the wide-lane scenario was created using real geographic data from Beijing Chang'an Avenue, representing a wide urban roadway.
Compared with the wide-lane scenario, the crossroad scenario serves as a relatively simple environment, featuring moderate building density and limited traffic flow, which facilitates the analysis of basic electromagnetic propagation behaviors. 
In contrast, the wide-lane scenario reflects a more complex and dynamic urban setting with heavy traffic, densely distributed roadside structures, and intricate spatial layouts, capturing the diversity and variability of real metropolitan environments. 
Together, these two scenarios represent the majority of practical UAV applications in urban areas, ranging from structured crossroads to large-scale arterial roads, and thus provide a comprehensive foundation for cross-scenario generalization analysis.

First, physical environments for the two scenarios are precisely rendered in AirSim. 
Subsequently, the corresponding 3D models in AirSim are properly imported into Wireless InSite to achieve geometric and positional alignment.
As a result, the precisely aligned electromagnetic spaces corresponding to the physical environments are established in Wireless InSite.
The UAV flight altitude is considered a key adaptive factor in practical UAV applications, as different missions and environments often require distinct operating heights to balance sensing coverage and communication performance.
In the crossroad scenario, the altitudes are set to 50 m, 70 m, and 80 m, representing typical low- and medium-altitude UAV operations in structured intersections.
In the wide-lane scenario, the altitudes are set to 200 m, 250 m, and 300 m, corresponding to higher-altitude flights suitable for broad-area observation and communication coverage over large arterial roads.
All surface materials, including buildings and roads, are set to concrete to ensure electromagnetic consistency between platforms.
To accommodate diverse communication application conditions, four carrier frequencies, i.e., 1.5 GHz, 5.9 GHz, 15 GHz, and 28 GHz, are configured, covering sub-6 GHz and mmWave bands commonly adopted in current and future UAV communication systems.
Specifically, twelve datasets are constructed under different environmental and communication conditions, as summarized in Table~\ref{twelve}.
\begin{table}[t]
\caption{Overview of the Twelve Datasets Under Different Environmental and Communication Conditions}
\centering
\renewcommand\arraystretch{1}
\setlength{\tabcolsep}{4pt} 
\label{twelve}
\begin{tabular}{c|ccccc}
\toprule
\textbf{Dataset} & \textbf{Scenario}  & \textbf{\makecell{Frequency\\(GHz)}} & \textbf{\makecell{Altitude\\(m)}}& \textbf{Snapshot}& \textbf{\makecell{Link\\(M)}} \\ 
\midrule
D1      & {\color{cyan}Crossroad} & {\color{violet}28}        & 50 &1830& 1.647      \\ \midrule
D2      & {\color{cyan}Crossroad} & {\color{violet}28}        & {\color{magenta}70}&1830&4.575      \\\midrule
D3      & {\color{cyan}Crossroad} & {\color{violet}28}        & 80&1830 &6.588       \\\midrule
D4      & {\color{cyan}Crossroad} & 15        & {\color{magenta}70} &1830 &4.575      \\\midrule
D5      & {\color{cyan}Crossroad} & 5.9       & {\color{magenta}70} &1830 &4.575     \\\midrule
D6      & {\color{cyan}Crossroad} & 1.6       & {\color{magenta}70}&1830 &4.575      \\\midrule
D7      & {\color{orange}Wide-lane}  & {\color{violet}28}        & 200&2090&13.376    \\\midrule
D8      & {\color{orange}Wide-lane}  & {\color{violet}28}        & {\color{magenta}250} &2090 &16.929     \\\midrule
D9      & {\color{orange}Wide-lane}  & {\color{violet}28}        & 300 &2090&20.900     \\\midrule
D10     & {\color{orange}Wide-lane}  & 15        & {\color{magenta}250}  &2090 &16.929    \\\midrule
D11     & {\color{orange}Wide-lane}  & 5.9       & {\color{magenta}250} &2090 &16.929    \\ \midrule
D12     & {\color{orange}Wide-lane}  & 1.6       & {\color{magenta}250} &2090 &16.929   \\ \midrule 
Sum     & -  & -      & - &23520 &128.527\\\bottomrule
\end{tabular}
\end{table}
These configurations collectively enable the collection of multi-condition sensing and channel multipath data across a wide range of altitudes, frequencies, and propagation environments.

\subsection{Data Collection}

To construct the dataset, AirSim is employed to capture RGB images, while Wireless InSite is used to obtain the U2G multipath parameters.
In AirSim, an RGB camera mounted beneath the UAV captures overhead images of the scenarios. 
In Wireless InSite, a transmit antenna is placed at the same position, while a ground antenna grid acts as a dense set of receiver (Rx) points. 
The coverage area of the ground antenna grid is designed to be identical to the field of view of the RGB image captured by the UAV, ensuring that the sensing range and the communication simulation area are perfectly aligned.
The dense antenna grid allows one-shot multipath acquisition for all Rx positions within the sensing area.

\begin{table}[t]
\caption{Trajectories of UAVs in Crossroad}
\centering
\renewcommand\arraystretch{1}
\setlength{\tabcolsep}{6pt} 
\label{t1}
\begin{tabular}{c|ccc}
\toprule
\textbf{Trajectory} & \textbf{\makecell{Movement velocity\\(m/snapshot)}}  & \textbf{\makecell{Start point\\(m)}} & \textbf{\makecell{End point\\(m)}} \\ 
\midrule
T1   & (0, -0.5) & (53.04, 87.25) & (53.04, -87.25)     \\ \midrule
T2   & (0, -0.5)   & (-2.25, 87.25) & (-2.25, -87.25)   \\\midrule
T3   & (0, -0.5)   & (-52.25, 87.25) & (-52.25, -87.25)\\\midrule
T4   & (-0.5, 0)   & (64.75, 33.67) & (-64.75, 33.67)\\\midrule
T5   & (-0.5, 0)   & (64.75, -4.33) & (-64.75, -4.33) \\\midrule
T6  & (-0.5, 0)    & (64.75, -42.33) & (-64.75, -42.33)\\\midrule
T7  & (-0.5, 0)    & (64.75, -80.33)  & (-64.75, -80.33)\\\bottomrule
\end{tabular}
\end{table}
\begin{table}[t]
\caption{Trajectories of UAVs in Wide-Lane}
\centering
\renewcommand\arraystretch{1}
\setlength{\tabcolsep}{8pt} 
\label{t2}
\begin{tabular}{c|ccc}
\toprule
\textbf{Trajectory} & \textbf{\makecell{Movement velocity\\(m/snapshot)}}  & \textbf{\makecell{Start point\\(m)}} & \textbf{\makecell{End point\\(m)}} \\ 
\midrule
T1   & (0, 8) & (-950, -1030) & (-950, 962)     \\ \midrule
T2   & (0, 8)   & (-300, -1030) & (-300, 962)   \\\midrule
T3   & (0, 8)   & (220, -1030) & (220, 962)\\\midrule
T4   & (0, 8)   & (690, -1030) & (690, 962)\\\midrule
T5   & (8, 0)   & (-864, -900) & (872, -900)\\\midrule
T6   & (8, 0)   & (-864, -600) & (872, -600) \\\midrule
T7  & (8, 0)    & (-864, -180) & (872, -180)\\\midrule
T8  & (8, 0)    & (-864, 460) & (872, 460)\\\midrule
T9  & (8, 0)    & (-864, 900)  & (872, 900)\\\bottomrule
\end{tabular}
\end{table}
To support massive data collection under dynamic UAV motion, flight trajectories are designed for batch scenario generation. 
The trajectories of UAVs in crossroad and wide-lane are presented in Tables~\ref{t1} and \ref{t2}, respectively.
Positions of UAV equipped with the camera and antenna are updated snapshot by snapshot, ensuring synchronization between AirSim and Wireless InSite. 
Internal configuration files in Wireless InSite are programmatically modified to emulate the synchronous movement of the transmit antenna and the ground antenna grid across snapshots.
The UAV conducted horizontal flights at a constant speed, following identical x-y trajectories at different altitudes with variations only in height.
This configuration ensures spatial consistency among data collected across altitudes and maintains precise correspondence between visual snapshots in AirSim and propagation simulations in Wireless InSite, thereby providing a robust foundation for cross-modal mapping from sensing to multipath parameters.

For each snapshot, two types of data are collected to establish a comprehensive dataset.
In AirSim, the RGB camera onboard the UAV captures RGB images that describe the visual and structural characteristics of the environment.
In Wireless InSite, the transmit antenna mounted at the same UAV position communicates with each element in the ground antenna grid, generating corresponding multipath parameters of the U2G links, including the power, delay, AoD, and AoA of the dominant propagation paths.
For data scale, in the crossroad scenario, each condition, i.e., under a specific flight altitude and frequency configuration, contains 1,830 snapshots, while the urban wide-lane scenario includes 2,090 snapshots per condition.
In total, the dataset consists of 23.52\,k RGB images and 0.13 billion sets of multipath parameters for the UAV-to-ground links.
Together, these synchronized sensing and communication data provide the basis for exploring cross-modal mapping from sensing to multipath parameters.



\section{Wireless Channel Foundation Model for Multipath Generation}

\begin{figure*}[t]
\centering
\includegraphics[width=\textwidth]{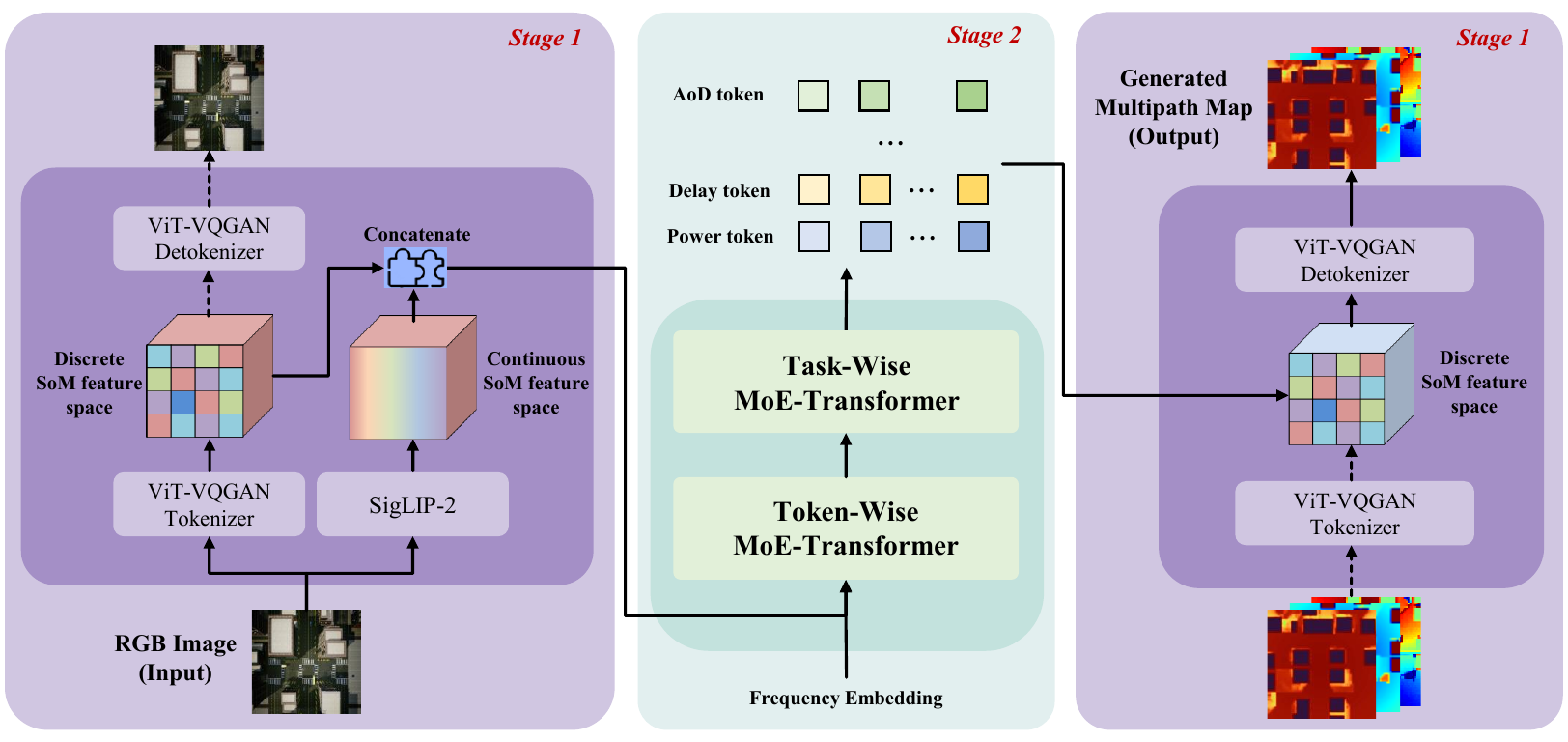}
\caption{An illustration of the network structure of the proposed WiCo-MG. The dashed arrows are only active during the ViT-VQGAN training in Stage~1.}
\label{fig1}
\end{figure*}

In this section, WiCo-MG is proposed, which is designed to generate multipath from RGB images in order to meet the massive and diverse data requirements of AI-native communication systems.
Cross-modal generation from RGB images to multipath maps is non-trivial due to two key challenges.
On the one hand, the representation discrepancy between sensing and electromagnetic domains is significant.
On the other hand, the mapping from sensing to multipath parameters is complex and frequency-dependent.
WiCo-MG follows a two-stage approach, as shown in Fig.~\ref{fig1}.
In stage 1, RGB images and multipath maps are embedded into the SoM feature space to form compact and well-aligned representations that facilitate both generation and semantic understanding.
In stage 2, S-R MoE Transformer is leveraged to achieve effective SoM feature space cross-modal mapping, enabling accurate and disentangled generation of multipath parameters with frequency-adaptive generalization.

\subsection{Stage 1: SoM Feature Space Embedding}

\subsubsection{RGB Image Representation Learning}

To effectively exploit sensing information from RGB images, both a discrete and a continuous SoM Feature Space are constructed for RGB features. 
The discrete SoM feature space represents images as quantized codes, which are compact and well-suited for generative modeling. 
The continuous SoM feature space preserves semantic richness and fine-grained relationships, supporting better understanding and alignment. 
Together, these two spaces enable the image features to possess both generative capability and semantic interpretability, thereby supporting more effective cross-modal alignment and representations.
Below, a detailed description of these two SoM feature spaces is provided.

\begin{enumerate}[label=\alph*)]
    \item \textbf{Discrete SoM Feature Space of RGB Image:} To begin with, the construction of the discrete SoM feature space is introduced. The vector-quantized variational autoencoder (VQVAE)~\cite{vqvae} learns discrete latent codes through a convolutional autoencoder, while VQGAN~\cite{vqgan} improves reconstruction quality by adding adversarial training and non-local attention. 
    ViT-VQGAN~\cite{vitvqgan} further replaces CNN-based encoders and decoders with ViTs as shown in Fig.~\ref{vitvqgan}, yielding more flexible representations and higher-fidelity reconstructions.
\begin{figure*}[t]
\centering
\includegraphics[width=6in]{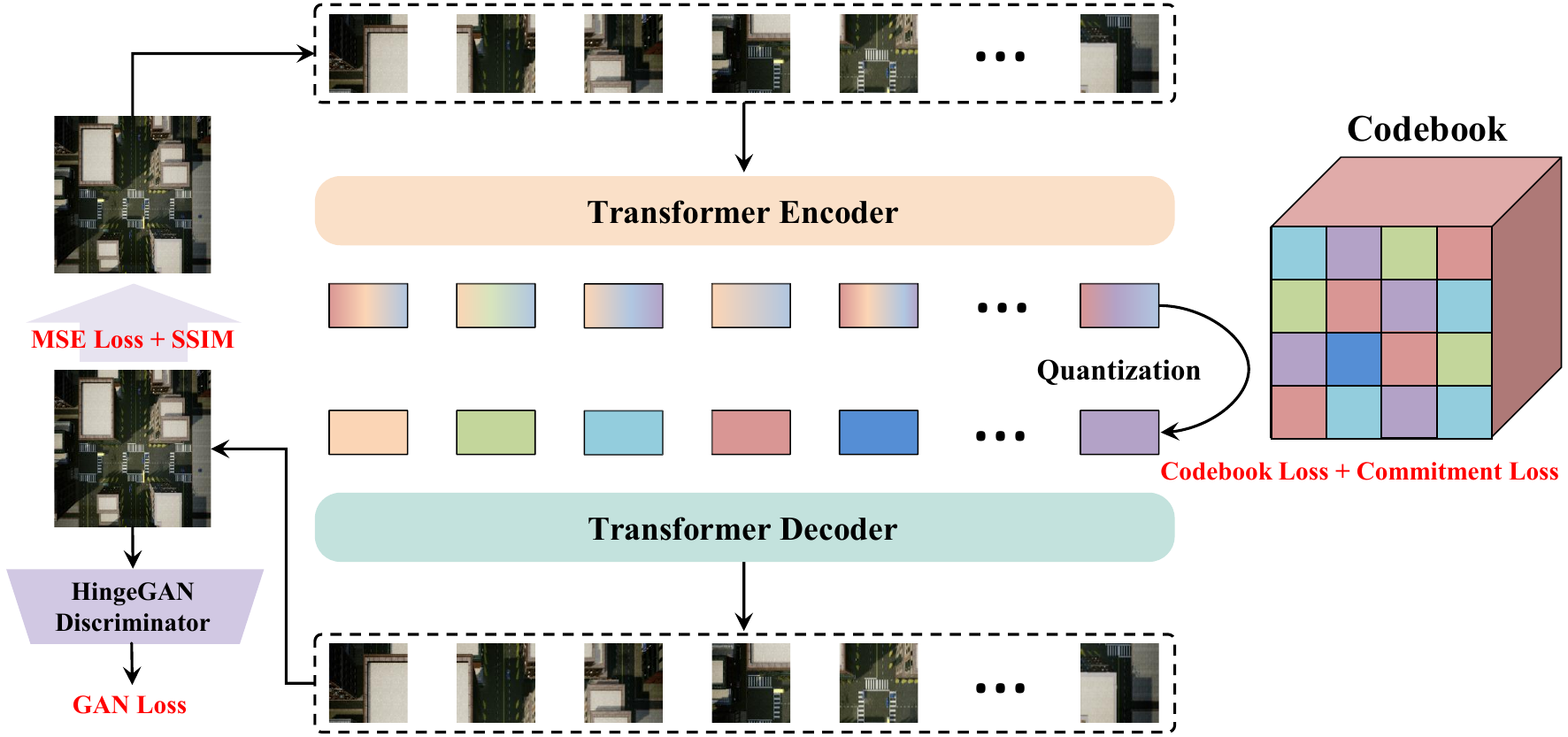}
\caption{An illustration of the network structure of ViT-VQGAN.}
\label{vitvqgan}
\end{figure*}
    Therefore, ViT-VQGAN is adopted as the encoder for the discrete SoM feature space, which provides compact codebook representations, thereby facilitating stable and effective cross-modal generation.

    Given an RGB image $I \in \mathbb{R}^{H_I \times W_I \times 3}$, a ViT-based encoder $E_I$ and a ViT-based decoder $D_I$ learn to represent it with codes from a learned and discrete codebook $\mathcal{Z}_I=\{z^I_k\}\subset \mathbb{R}^{n_z}$, where $n_z$ is the dimensionality of codes.
    Specifically, the encoder $E_I$ of the ViT-VQGAN first divides the image into non-overlapping patches of size $p_I \times p_I \times 3$, and then transforms these patches into image tokens.
    Image tokens are denoted as 
    \begin{equation}
        I_{\rm t}=E_I(I) \in \mathbb{R}^{h_I \times w_I \times n_z},
    \end{equation}
    where $h_I=H_I/p_I$ and $w_I=W_I/p_I$.
    Then, an element-wise quantization operator $\mathbf Q(\cdot)$ is adopted to map each image token to its closest codebook entry $z^I_k$, which is determined by
    \begin{equation}
    \begin{split}
        z^{I}_{ij}=\mathbf Q(I_{{\rm t}, ij})=(\arg\min_{z^I_s\in \mathcal{Z}} \|I_{{\rm t}, ij}- z^I_s\|)\in \mathbb{R}^{n_z},\\
        i=1,2,\dots,h_I, j=1,2,\dots,w_I.
    \end{split}
    \end{equation}
    Finally, the decoder $D_I$ of the ViT-VQGAN performs the inverse operation, mapping image tokens $z^I_{\rm q}=\{z^{I}_{ij}\}$ back to $p_I \times p_I \times 3$ image patches and regrouping them into an image $\hat I$, which is given by
    \begin{equation}
        \hat I=D_I(z^I_{\rm q})\in \mathbb{R}^{H_I \times W_I \times 3}.
    \end{equation}
    The non-differentiability of the quantization operation is handled using a straight-through gradient estimator, which directly passes the gradients from the decoder back to the encoder~\cite{tidu}. This allows both the model and the codebook to be optimized jointly in an end-to-end manner.

    To enable the joint optimization of both the encoder-decoder network and the codebook, a composite loss function is adopted, which is given by
    \begin{equation}\label{VQloss}
        L_{\rm VQ}=L_{\rm rec}+\|{\rm sg}(I_{\rm t})-z^I_{\rm q}\|_2^2+\beta \|{\rm sg}(z^I_{\rm q})-I_{\rm t}\|_2^2,
    \end{equation}
    where reconstruction loss $L_{\rm rec}=L_{\rm mse}+L_{\rm SSIM}$, ${\rm sg}(\cdot)$ represents the stop-gradient operation, $\|{\rm sg}(I_{\rm t})-z^I_{\rm q}\|_2^2$ is the codebook loss, and $\|{\rm sg}(z^I_{\rm q})-I_{\rm t}\|_2^2$ is the commitment loss with weighting factor $\beta$~\cite{vqvae}.
    For the reconstruction loss, the structural similarity index (SSIM) is incorporated to better preserve local spatial structures. 
    Compared with L1 or L2 alone, SSIM is more effective in capturing fine-grained details such as occlusions, edge discontinuities, and sharp transitions, thereby improving the fidelity of the reconstructed features.

    Furthermore, an adversarial mechanism based on the HingeGAN objective is introduced to enhance the perceptual realism of reconstructed outputs. 
    The discriminator is trained to distinguish real samples from reconstructed ones, while the generator learns to fool the discriminator $C_I$ under the hinge loss formulation, which is denoted as 
    \begin{equation}\label{GANloss}
        L_{\rm GAN}=\log C_I(I)+\log (1-C_I(\hat I)).
    \end{equation}
    Compared with conventional adversarial losses, HingeGAN provides more stable gradients and sharper reconstructions, thereby improving the fidelity of fine textures and local structures.

    Overall, the complete objective is given as
    \begin{equation}\label{totalloss}
    \begin{split}
        \mathcal{O}_I=\arg\min_{E_I,D_I,\mathcal{Z}_I}\max_{C_I} \mathbb{E}(L_{\rm VQ}(E_I,D_I,\mathcal{Z}_I)+\\\lambda L_{\rm GAN}(E_I,D_I,\mathcal{Z}_I,C_I)),
    \end{split}
    \end{equation}
    where adaptive weight $\lambda$ can be denoted by
    \begin{equation}
        \lambda=\frac{\nabla_l (L_{\rm rec})}{\nabla_l (L_{\rm GAN})+\delta},
    \end{equation}
    where $\nabla_l$ denotes the gradient of its input with respect to the last layer $l$ of the decoder and $\delta=10^{-6}$ is used for numerical stability.
    
    \item \textbf{Continuous SoM Feature Space of RGB Image:} 
    In addition to the discrete latent representation, a continuous SoM feature space is further constructed to preserve the semantic richness and global contextual information of RGB images. 
    To this end, we adopt SigLIP-2~\cite{siglip}, a large-scale vision-language pretrained encoder that improves upon CLIP by replacing contrastive training with a sigmoid-based loss function. 
    This training paradigm allows SigLIP-2 to align visual features with textual semantics more effectively, leading to embeddings that are both discriminative and semantically consistent. 
    Given an input image $I \in \mathbb{R}^{H \times W \times 3}$, SigLIP-2 projects it into a dense continuous embedding $s^I_c \in \mathbb{R}^{n_c \times d}$, where $n_c$ denotes the number of embedding vectors and $d$ is the embedding dimension. 
    Compared with discrete quantization, this continuous representation retains fine-grained semantics, global contextual cues, and cross-modal alignment capabilities, which are critical for channel feature generation. 
    By integrating SigLIP-2 embeddings, our model leverages semantically enriched features that complement the compact generative representation provided by the discrete latent space, resulting in a more robust and flexible cross-modal mapping.

\end{enumerate}

After constructing the discrete and continuous SoM feature spaces, a fusion mechanism is further designed to integrate them effectively. The discrete tokens, obtained from ViT-VQGAN quantization, serve as the primary representation for generative modeling, while the continuous embeddings from SigLIP-2 provide complementary semantic guidance. To combine them, a gated modulation strategy is adopted. 
The continuous embeddings are projected and aggregated into a global conditioning vector, which is then broadcast to match the length of the discrete token sequence. 
A gating function, computed from the discrete tokens via a sigmoid activation, adaptively controls how much semantic information from the continuous space is injected. 
The final token representation is obtained as
\begin{equation}
    \tilde{z} = z^{I'}_{\rm q} + \sigma\!\big(g_c(z^{I'}_{\rm q})\big) \odot h(s^I_{\rm c})\,\alpha \in \mathbb{R}^{h_I \times w_I \times d},
\end{equation}
where $z^{I'}_{\rm q}=L(z^{I}_{\rm q})$ denotes the embedded discrete tokens through a linear layer, $h(s^I_{\rm c})$ is the projected conditioning from continuous embeddings, $g_c(\cdot)$ is the gating function, $\sigma(\cdot)$ is the sigmoid function, $\odot$ denotes element-wise multiplication, and $\alpha$ is a scaling factor to regulate the influence.
This fusion strategy allows the model to retain the compact and generative properties of discrete tokens, while flexibly leveraging semantic cues from the continuous space for improved cross-modal alignment.

\subsubsection{Multipath Map Representation Learning}
Each physical parameter of the multipath channel, such as power, delay, AoD, and AoA, is organized into a separate SoM feature space to ensure parameter-wise disentanglement. 
We focus on the discrete SoM feature space, where each multipath map is embedded using a ViT-VQGAN encoder into compact quantized tokens. 

Similarly, given a multipath map $M \in \mathbb{R}^{H_M \times W_M \times 1}$ related to one of the multipath parameters, a ViT-based encoder $E_M$ and a ViT-based decoder $D_M$ learn to represent it with codes from a discrete codebook $\mathcal{Z}_M=\{z^M_k\}\subset \mathbb{R}^{n_z}$, where $n_z$ is the dimensionality of codes.
It is worth noting that the codebook $\mathcal{Z}_M$ is initialized from the optimal codebook learned in the RGB image domain. 
This initialization strategy narrows the gap between sensing and channel representations, providing a shared basis across modalities. 
As a result, the model benefits from faster convergence and improved cross-modal alignment.
Then, the encoder $E_M$ of the ViT-VQGAN first divides the multipath map into non-overlapping patches of size $p_M \times p_M \times 3$, and then transforms these patches into multipath map tokens.
Multipath map tokens are denoted as 
\begin{equation}
    M_{\rm t}=E_M(M) \in \mathbb{R}^{h_M \times w_M \times n_z},
\end{equation}
where $h_M=H_M/p_M$ and $w_M=W_M/p_M$.
Next, an element-wise quantization operator $\mathbf Q(\cdot)$ is adopted to map each image token to its closest codebook entry $z^M_k$, which is determined by
\begin{equation}
\begin{split}
    z^{M}_{ij}=\mathbf Q(M_{{\rm t}, ij})=(\arg\min_{z^M_s\in \mathcal{Z}} \|M_{{\rm t}, ij}- z^M_s\|)\in \mathbb{R}^{n_z},\\
    i=1,2,\dots,h_M, j=1,2,\dots,w_M.
\end{split}
\end{equation}
Finally, the decoder $D_M$ of the ViT-VQGAN performs the inverse operation, mapping multipath map tokens $z^M_{\rm q}=\{z^{M}_{ij}\}$ back to $p_M \times p_M \times 3$ multipath map patches and regrouping them into an multipath map $\hat M$, which is given by
\begin{equation}
    \hat M=D_M(z^M_{\rm q})\in \mathbb{R}^{H_M \times W_M \times 3}.
\end{equation}
The subsequent adversarial training, loss function, and overall optimization objective follow the same procedure as the ViT-VQGAN framework used for RGB images. 
In particular, the combination of reconstruction, commitment, and adversarial losses is employed without modification, ensuring consistency in training across both perception and channel domains.

\subsection{Stage 2: SoM Feature Space Mapping}
In this stage, the primary objective is to establish a mapping between the sensing and channel SoM feature spaces. 
Through this process, the representations extracted from an RGB image are transformed into corresponding multipath parameter maps, including power, delay, and angle. 
This mapping is implemented by an S-R MoE Transformer, in which both token-wise and task-wise MoE mechanisms are employed to achieve a balance between shared knowledge and task-specific adaptation. 
In addition, communication frequency embeddings are incorporated into the gating network, enabling frequency-aware adaptation and enhancing the model's generalization capability across different frequency bands.
The token-wise MoE and task-wise MoE are described as follows.

\subsubsection{Token-Wise MoE}
\begin{figure}[t]
\centering
\includegraphics[width=3.5in]{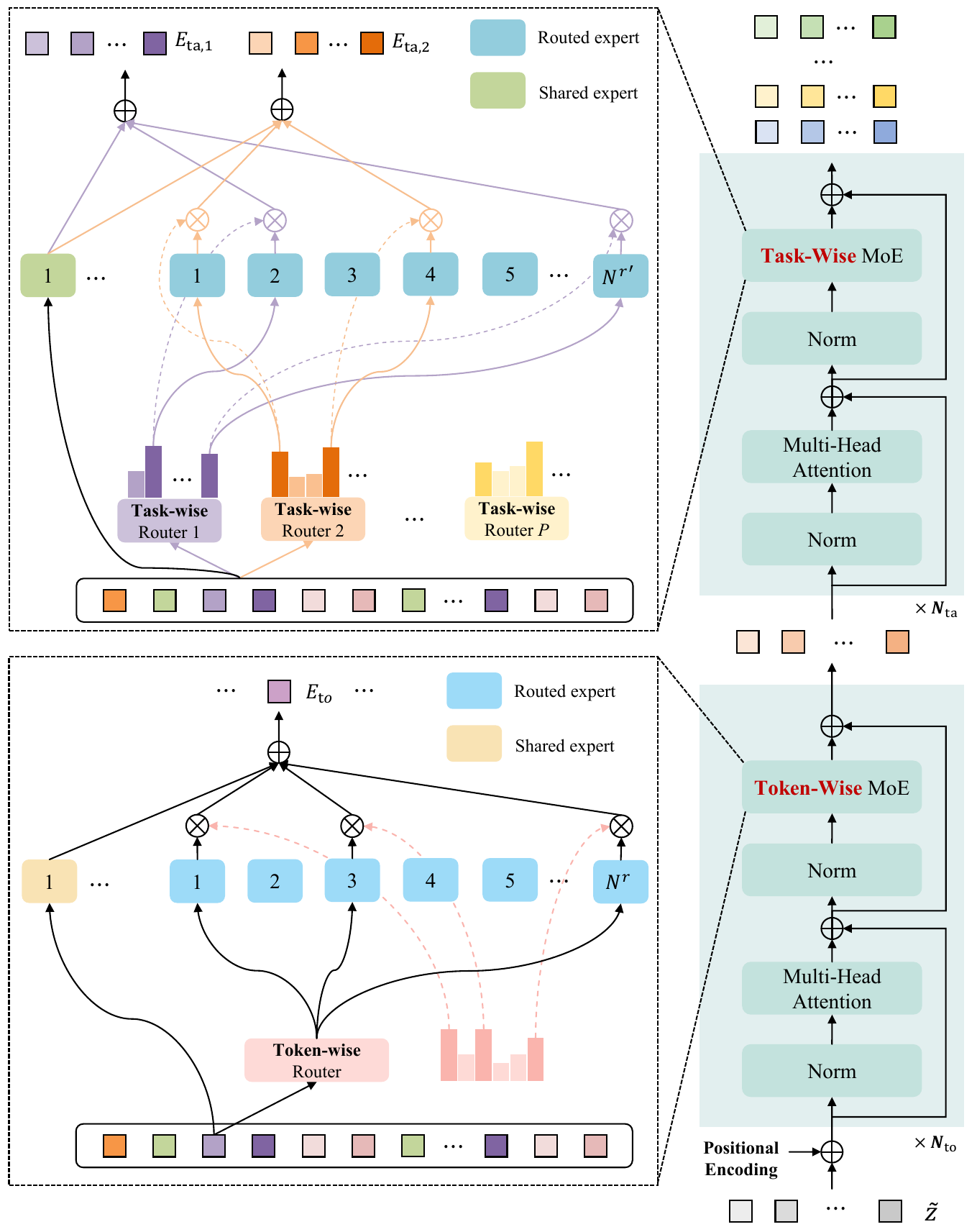}
\caption{An illustration of the structure of token-wise MoE Transformer and task-wise MoE Transformer.}
\label{tokenmoe}
\end{figure}

The token-wise MoE mechanism is designed to enable fine-grained expert activation at the token level, following the routing strategy introduced in DeepSeek-V3~\cite{deepseek}. 
Instead of assigning the same set of experts to all tokens within a layer, each token is dynamically routed to its most relevant experts according to learned gating scores, as shown in Fig.~\ref{tokenmoe}.
Meanwhile, compared with conventional MoE architectures, token-wise MoE uses finer-grained experts and isolates some experts as shared ones.
To be specific, for each input image token $\tilde{z}_{ij}\in \mathbb{R}^{d}$, the output of token-wise MoE layer is computed as 
\begin{equation}
    E_{\rm to}=\sum_{k=1}^{N_{\rm s}} {\rm FFN}_k^{\rm s}(\tilde{z}_{ij})+
    \sum_{k=1}^{N_{\rm r}} g_k{\rm FFN}_k^{\rm r}(\tilde{z}_{ij}),
\end{equation}
where
\begin{equation}
    g_k=
    \begin{cases}
    \hat{g}_k, & \hat{g}_k \in \mathrm{Topn}\big(\{\hat{g}_l \mid 1 \le l \le N_r\}, K_r\big), \\
    0, & \text{otherwise,}
    \end{cases}
\end{equation}
\begin{equation}
    \hat{g}_k={\rm softmax}({\rm g}_f({\rm g}_t(\tilde{z}_{ij})+E_{\rm f})),
\end{equation}
where $N_{\rm s}$ and $N_{\rm r}$ denote the numbers of shared experts and routed experts, respectively; 
${\rm FFN}_k^{\rm s}(\cdot)$ and ${\rm FFN}_k^{\rm r}(\cdot)$ denote the $k$-th shared experts and the $k$-th routed experts, respectively;
$\hat{g}_k$ is the gating value for the $k$-th routed experts and $g_k$ is the Top-$K_r$ results;
${\rm g}_t(\cdot)$ is the gating network of the image token $\tilde{z}_{ij}$;
$E_{\rm f}$ represents the communication frequency embedding;
${\rm g}_f(\cdot)$ is the gating network of the frequency-embedded image token;
${\rm softmax}(\cdot)$ represents the softmax function.
Note that $\tilde{z}_{ij}$ not only refers to the token obtained from the Stage 1 but also includes the intermediate token representation output by the preceding Transformer blocks in Stage 2.
After the computation process defined above, the functions of each component can be further clarified.
\begin{enumerate}[label=\alph*)]
\item
First, the token-wise MoE incorporates both shared experts and routed experts. 
The shared experts capture global representations that are beneficial across all tokens and tasks, thereby enhancing model stability and preventing overfitting. 
In contrast, routed experts are dynamically activated based on token-dependent gating scores, allowing the model to specialize in localized and task-specific features. 
This hybrid design ensures a balance between global generalization and local specialization.
\item
Then, a frequency-aware gating mechanism is introduced.
Specifically, the communication frequency is embedded into the routing network through two linear layers, which can be denoted by
\begin{equation}
    E_{\rm f}={\rm Emb}(f)\in \mathbb{R}^{d},
\end{equation}
where ${\rm Emb}(\cdot)$ represents the communication frequency embedding network.
This design enables the router to adjust expert selection adaptively under different frequency bands, making the model more robust to spectral variations in wireless channels.
\item
Next, a Top-$K_r$ routing strategy is adopted, where only the routed experts with the highest gating scores are activated for each token. 
This sparse activation significantly reduces computational overhead while maintaining the representational diversity required for complex cross-modal mappings.

\end{enumerate}
Overall, the token-wise MoE layer aggregates the outputs of both shared and routed experts weighted by their gating scores, enabling adaptive and frequency-aware feature transformation between the perception and channel SoM feature spaces.

\subsubsection{Task-Wise MoE}
The task-wise MoE mechanism is introduced to achieve parameter-level specialization across different multipath generation tasks, such as power, delay, and angle.  
Unlike the token-wise MoE, where each token is dynamically routed to distinct experts, the task-wise MoE assigns the same set of experts to all tokens within a task, while different tasks are routed to different expert subsets.
The structure of the task-wise MoE Transformer is shown in Fig.~\ref{tokenmoe}.
In this way, each task learns from a dedicated combination of shared and routed experts, enabling consistent token processing within a task and promoting specialization across heterogeneous multipath parameters. 
For input image tokens $\tilde{z}\in \mathbb{R}^{h_I \times w_I \times d}$, the output of task-wise MoE layer for task $p$ is computed as 
\begin{equation}
    E_{{\rm ta},p}=\sum_{k=1}^{N^{'}_{\rm s}} {\rm FFN}_k^{\rm s'}(\tilde{z})+
    \sum_{k=1}^{N^{'}_{\rm r}} g^{'}_{k,p}{\rm FFN}_k^{\rm r'}(\tilde{z}),
\end{equation}
where
\begin{equation}
    g^{'}_{k,p}=
    \begin{cases}
    \hat{g}^{'}_{k,p}, & \hat{g}^{'}_{k,p} \in \mathrm{Topn}\big(\{\hat{g}^{'}_{l,p} \mid 1 \le l \le N^{'}_r\}, K^{'}_r\big), \\
    0, & \text{otherwise,}
    \end{cases}
\end{equation}
\begin{equation}
    \hat{g}^{'}_{k,p}={\rm softmax}({\rm g}^{'}_{t,p}(\tilde{z})),
\end{equation}
where $N^{'}_{\rm s}$ and $N^{'}_{\rm r}$ denote the numbers of shared experts and routed experts, respectively; 
${\rm FFN}_k^{\rm s'}(\cdot)$ and ${\rm FFN}_k^{\rm r'}(\cdot)$ denote the $k$-th shared experts and the $k$-th routed experts, respectively;
$\hat{g}^{'}_{k,p}$ is the gating value for the $k$-th routed experts and $g^{'}_{k,p}$ is the Top-$K^{'}_r$ results;
${\rm g}^{'}_{t,p}(\cdot)$ is the gating network of image tokens $\tilde{z}$;
${\rm softmax}(\cdot)$ represents the softmax function.
Note that $\tilde{z}$ not only refers to the tokens obtained from the Stage~1 but also includes the intermediate token representations output by the preceding Transformer blocks in Stage~2.
After the computation defined above, the roles of the experts and the gating strategy can be clarified.
The shared experts ${\rm FFN}_k^{\rm s'}(\cdot)$ provide task-invariant representations that capture common structural patterns across all multipath parameters, while the routed experts ${\rm FFN}_k^{\rm r'}(\cdot)$ specialize in parameter-dependent modeling guided by task-specific gating networks.
In this design, each output task is equipped with an independent gating network ${\rm g}^{'}_{t,p}(\cdot)$, which assigns the same subset of experts to all tokens within that task.
Furthermore, only the Top-$K^{'}_r$ routed experts with the highest gating scores are activated, forming a compact and specialized expert subset that improves both computational efficiency and task-level adaptability.

To integrate the two MoE mechanisms within a unified framework, a segmented stacking architecture is employed, in which each Transformer encoder block incorporates a MoE layer as its feed-forward layer.
In this configuration, $N_{\rm to}$ blocks equipped with token-wise MoE layers are first applied to refine the sensing representations extracted from RGB images, enabling fine-grained expert activation and comprehensive modeling of spatial and semantic dependencies.
Subsequently, $N_{\rm ta}$ blocks with task-wise MoE layers are stacked on top to inject task-specific signals and achieve parameter-level specialization for multipath generation.
This segmented stacking strategy allows the network to progressively evolve from sensing-driven feature refinement to task-oriented multipath parameter generation, thereby establishing a coherent mapping from sensing to multipath parameters.
Then, the output of the last Transformer block is processed by the corresponding ViT-VQGAN decoder $D_{M,p}$ to obtain the generated multipath map $\hat{M}_p\in \mathbb{R}^{H_M \times W_M \times 1}$.

\section{Experiments}

In this section, the experiment settings are illustrated, and then the performance of the proposed WiCo-MG is evaluated.
{\color{black}Specifically, the in-distribution generation ability and out-of-distribution generalization ability of WiCo-MG are evaluated.
Then, the ablation study is conducted to investigate the contributions of each component.
Next, scaling analysis and extensibility analysis are conducted.
Finally, the network storage and inference cost of the WiCo-MG and baselines are evaluated.}

\subsection{Experiment Settings}

\subsubsection{Dataset Overview}
Twelve datasets are utilized to comprehensively evaluate the capability of the proposed WiCo-MG.
Each dataset corresponds to a unique combination of scenario, carrier frequency, and UAV flight altitude.
{\color{black}Each sample is composed of one RGB image from UAV, a corresponding communication frequency indicator, and six multipath parameter maps related to the first dominant path of the U2G link.}
These maps respectively represent power, delay, AoD (azimuth and elevation), and AoA (azimuth and elevation).
It is worth noting that, except for the experiments presented in \ref{NewMultipath}, all other experiments are conducted using the four maps of power, delay, and AoD (azimuth and elevation).

\subsubsection{Baseline}
For a comprehensive comparison, the following three baselines are provided, including Transformer, ResNet, and advanced LLM-based methods.
{\color{black}Specifically, the Transformer and ResNet serve as references for evaluating full-sample learning performance, whereas the LLM-based methods are employed to assess the generalization capability of the proposed WiCo-MG.}
\begin{enumerate}[label=\alph*)]
    \item \textbf{Transformer \cite{transformer}}: A ViT~\cite{vit}-style patching scheme is adopted, where the RGB image is divided into non-overlapping patches and linearly projected to form image tokens. The tokens are processed by a standard Transformer, and the output token sequence is decoded by a lightweight decoder to produce the multi-channel multipath parameter maps. For a fair comparison, the patching is kept consistent with WiCo-MG, and identical data splits and evaluation metrics are used.
    \item \textbf{ResNet \cite{resnet}}: A convolutional baseline is implemented by feeding the RGB image directly into a ResNet-style model that transforms images to the multi-channel multipath parameter maps. This baseline assesses the effectiveness of hierarchical convolutional features for the cross-modal generation task under the same evaluation settings.
    \item \textbf{LLM-Based Model \cite{GPT}}: A GPT-2-based model is employed to examine the capability of LLMs for cross-modal generation. Pretrained GPT-2 weights are loaded, and only a small subset of layers is fine-tuned. The RGB image is tokenized via the same ViT-style patching. The resulting tokens are fed into GPT-2, and the output tokens are transformed by several convolutional and linear layers to yield the multipath parameter maps.
\end{enumerate}

\subsubsection{Network and Training Configuration}
\begin{table*}[t]
\caption{Network Parameters of WiCo-MG}
\centering
\renewcommand\arraystretch{1}
\label{parameter1}
\begin{tabular}{c|ccccccc}
\toprule 
\textbf{Model}         & \textbf{S1.depth} & \textbf{S1.width} & \textbf{Codebook} & \textbf{S2.depth} & \textbf{S2.width} & \textbf{Num.expert} & \textbf{Parameters (M)} \\ \midrule
WiCo-MG-Small & 4        & 256      & 64      & 3+1        & 128      & 2/5        &  13.15M/24.19M             \\ \midrule
WiCo-MG-Base  & 8        & 512      & 128      & 6+2        & 256      & 3/9        &  66.64M/137.90M     \\ \midrule
WiCo-MG-Large & 16        & 1024      & 512      & 12+4        & 512      & 5/20        &   418.72M/887.52M    \\ \bottomrule 
\end{tabular}
\end{table*}
{\color{black}To demonstrate the scaling ability of WiCo-MG,} Table \ref{parameter1} presents the network parameters for the WiCo-MG, detailing 3 different sizes to investigate the impact of model size on performance.
\begin{table}[t]
\centering
\renewcommand\arraystretch{1}
\caption{Hyper-Parameter for Network Training}
\label{parameter2}
\begin{tabular}{c|c}
\toprule 
\textbf{Parameter  }       & \textbf{Value} \\ \midrule
Batch size      &  64     \\ \midrule
Starting learning rate  &  Stage 1: $2\times 10 ^{-4}$; Stage 2: $4.5\times 10 ^{-4}$      \\ \midrule
Learning rate scheduler &   ReduceLROnPlateau (factor=0.5, \\&patience=10, min\_lr=$1\times10^{-6}$)    \\ \midrule
Epochs &   500  \\ \midrule
Optimizer & ADAM      \\
\bottomrule 
\end{tabular}
\end{table}
Table \ref{parameter2} outlines the hyper-parameters used for training the WiCo-MG. In the experiments, the size of each patch is set as (8, 8).
The network is trained using PyTorch with the Adaptive Moment Estimation (ADAM) optimizer~\cite{diederik2014adam}.
All experiments are conducted on the same machine with an NVIDIA RTX 4090 GPU and an Intel Xeon Platinum 8358P CPU.

\subsubsection{Loss Function and Performance Metric}
In Stage 1, the loss function has been defined as (\ref{VQloss}), (\ref{GANloss}), and (\ref{totalloss}).
In Stage 2, the outputs are multipath maps $\hat{M}_p\in \mathbb{R}^{H_M \times W_M \times 1}, p=1,2,\dots,P$, where $P$ is the number of generated multipath parameters. The loss function and performance metric of each multipath map are based on the normalized mean squared error (NMSE), which can be calculated as
\begin{equation}
    NMSE(M_p,\hat{M}_p)=
    \frac{\|M_p-\hat{M}_p\|^2}{\|\hat{M}_p\|^2}, p=1,2,\dots,P,
\end{equation}
where the NMSE is computed element-wise over all entries.
To address the imbalance among different multipath parameter generation tasks, a dynamic weight averaging (DWA) strategy~\cite{dwa} is employed to adaptively adjust the loss weights during training.
Specifically, the relative rate of loss reduction for each task is monitored over consecutive epochs, and tasks with slower convergence, i.e., harder to optimize, are assigned higher weights.
This mechanism encourages the model to focus more on difficult tasks while preventing overfitting to those that are easier to learn.
In our setting, DWA is applied across several multipath parameter losses.
Although these parameters are normalized to comparable numerical ranges, their semantic and physical impacts differ significantly. 
The utilization of DWA thus ensures balanced optimization and interpretable cross-modal learning behavior.



\subsection{In-Distribution Generation Ability}

\subsubsection{Performance Under Single-Scenario Conditions}
\begin{figure}[t]
\centering
\includegraphics[width=3.5in]{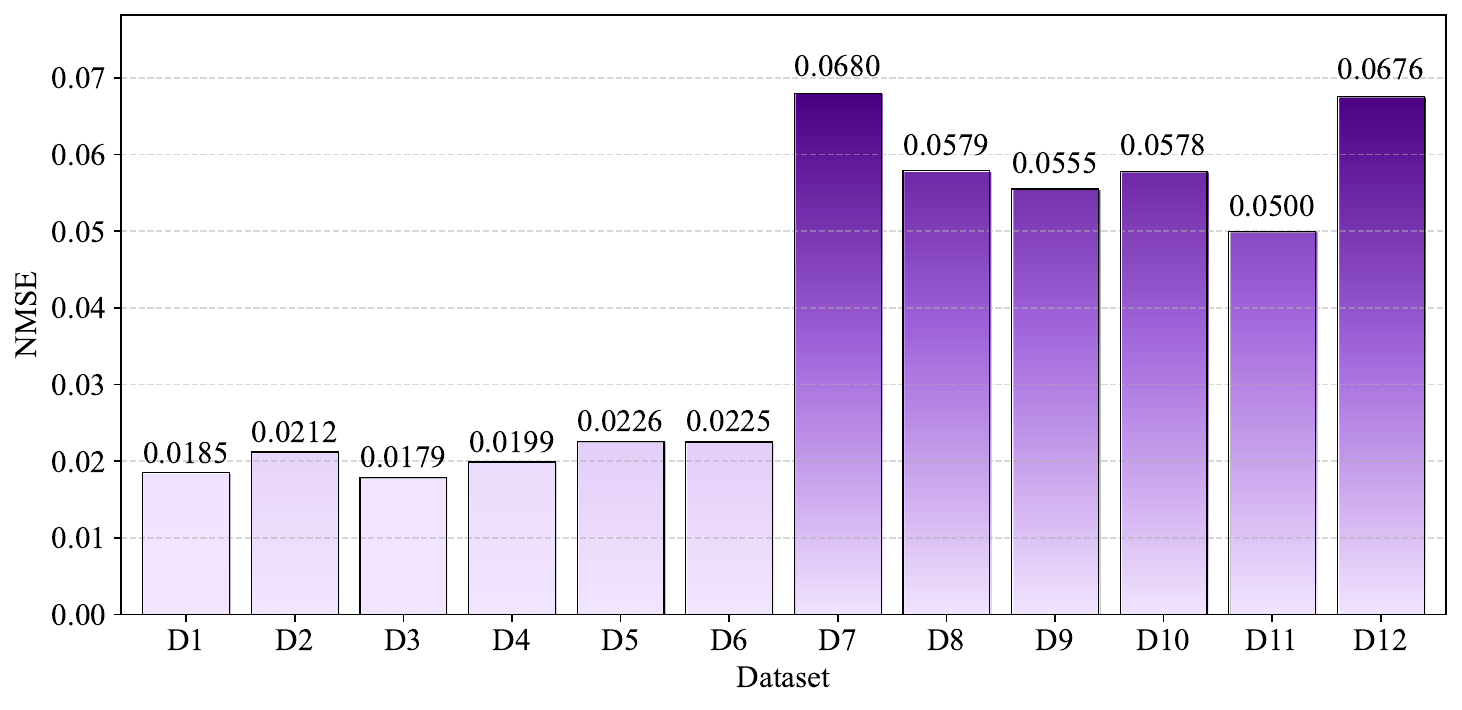}
\caption{Performance of WiCo-MG-Base under single-scenario conditions.}
\label{single}
\end{figure}

To evaluate the in-distribution generation ability of the proposed WiCo-MG, experiments are first conducted under single-scenario conditions, where training and testing are performed on data from the same scenario and communication configuration.
Each dataset listed in Table~\ref{twelve} (D1-D12) is independently used to train and evaluate the model.
This setting allows a direct comparison of the model's performance across diverse spatial layouts, frequency bands, and flight altitudes without cross-domain interference.
Under single-scenario conditions, the proposed model achieves consistently low generation errors across all datasets with NMSE values below 0.07, as shown in Fig. \ref{single}.
Moreover, WiCo-MG achieves an NMSE reduction of 6.33~dB compared with the baseline results in Table~\ref{Multiple}.
Performance differences are mainly observed across scenarios.
In the crossroad scenario, the model attains slightly lower NMSE due to the relatively regular spatial structure and fewer dynamic scatterers, while in the wide-lane scenario, the increased openness and richer multipath components lead to marginally higher generation errors.
By comparing the results under different UAV flight altitudes in the wide-lane scenario, it can be observed that higher altitudes enable the UAV to capture a broader environmental view and acquire more comprehensive spatial information, which leads to lower NMSE values and better generation performance.

\subsubsection{Performance Under Multiple-Scenario Conditions} 
{\color{black}Real-world environments and communication conditions are inherently diverse, necessitating models that can perform reliably across diverse environments.}
For this reason, the performance of WiCo-MG is evaluated under multiple-scenario conditions, where all datasets (D1-D12) are jointly used for training and testing.
Specifically, the samples from all datasets are merged, randomly shuffled, and then divided into training and testing sets, ensuring that both sets contain data from diverse scenarios, frequencies, and flight altitudes.
This joint training-testing configuration allows the model to learn from a richer and more heterogeneous data distribution, where information from different environments can complement and reinforce each other.
\begin{table}[t]
\caption{Performance of WiCo-MG-Base Under Multiple-Scenario Conditions and Performance of Dataset-Tailored Baselines}
\centering
\renewcommand\arraystretch{1}
\setlength{\tabcolsep}{9pt} 
\label{Multiple}
\begin{tabular}{c|ccc}
\toprule 
\textbf{Dataset}  & \textbf{WiCo-MG-Base} & \textbf{Transformer} & \textbf{ResNet} \\ \midrule   
D1     & 0.0336 & 0.1187 & 0.0791 \\\midrule 
D2      & 0.0109 & 0.1259 & 0.0869 \\\midrule
D3       & 0.0154 & 0.1145 & 0.0932 \\\midrule
D4        & 0.0110 & 0.1059 & 0.0832 \\\midrule
D5     & 0.0106 & 0.1023 & 0.0809 \\\midrule
D6       & 0.0106 & 0.1001 & 0.0797 \\\midrule
D7      & 0.0535 & 0.3185 & 0.2537 \\\midrule
D8      & 0.0476 & 0.3415 & 0.2663 \\\midrule
D9    & 0.0597 & 0.3679 & 0.2902 \\\midrule
D10   & 0.0481 & 0.3117 & 0.2360 \\\midrule
D11      & 0.0493 & 0.3275 & 0.2502 \\\midrule
D12      & 0.0513 & 0.3353 & 0.2578 \\\midrule
Average NMSE &  \textbf{0.0335} & \textbf{0.2225} & \textbf{0.1714} \\ \bottomrule
\end{tabular}
\end{table}
As shown in Table~\ref{Multiple}, the model trained under multiple-scenario conditions achieves consistently lower NMSE compared with single-scenario condition training, with the average NMSE reduced by approximately 0.77~dB.
These findings confirm that learning from diverse scenarios enhances the model's robustness and its ability to generalize across complex propagation environments.
Jointly trained WiCo-MG-Base surpasses the dataset-tailored baselines that are individually optimized for their respective scenario conditions, demonstrating an NMSE improvement of 7.09~dB over the baseline methods reported in Table~\ref{Multiple}.
This indicates that the learned representations enable a single model to be utilized across multiple environments without the need for retraining.

\subsection{Out-of-Distribution Generalization Ability}

\subsubsection{Performance Across Communication Frequency Bands}
The proposed WiCo-MG-Base model is pretrained on datasets collected at 28, 15, and 5.9~GHz, and subsequently evaluated on the unseen 1.6~GHz dataset to assess its generalization capability.
\begin{figure}[t]
\centering
\includegraphics[width=3.5in]{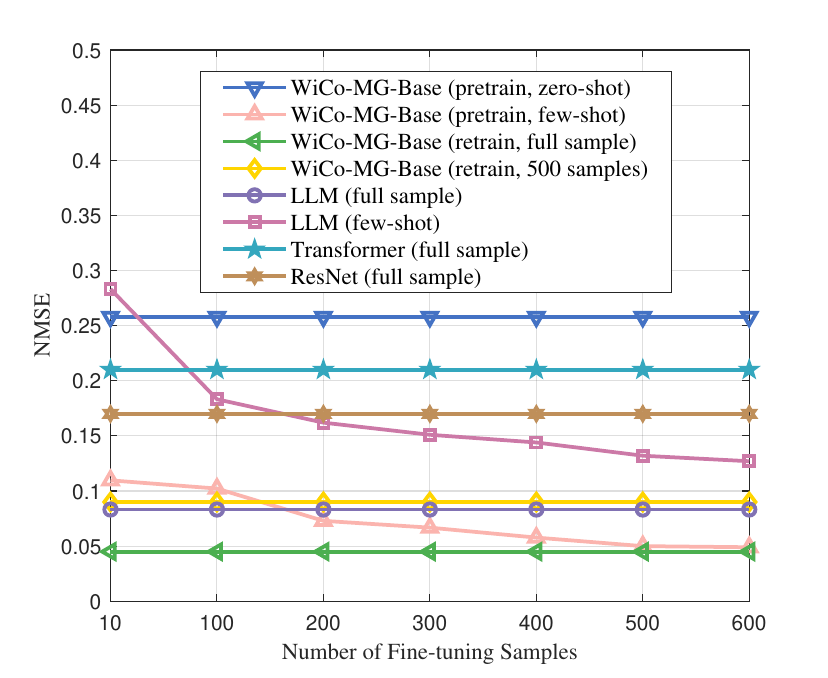}
\caption{Performance of WiCo-MG-Base and other baselines across communication frequency bands.}
\label{frequency}
\end{figure}
{\color{black}As shown in Fig.~\ref{frequency}, the few-shot learning results of the pretrained WiCo-MG-Base indicate that approximately 500 fine-tuning samples of the 1.6~GHz scenario are sufficient to achieve a performance gap of less than 0.01 in NMSE compared with full-sample retraining (with reference to pink triangle line and green left-pointing-triangle line).}
In contrast, direct retraining on 500 samples without pretraining fails to reach the same level of accuracy, which demonstrates the effectiveness of the cross-frequency pretraining (with reference to yellow diamond line and pink triangle line). 

Furthermore, with only 200 fine-tuning samples, the pretrained WiCo-MG-Base surpasses the fine-tuned LLM-based model on the full samples of the 1.6~GHz dataset (with reference to pink triangle line and purple circle line). 
All of the aforementioned results related to WiCo-MG-Base also exceed the performance of conventional deep learning baselines trained with full samples of the 1.6~GHz dataset, including Transformer and ResNet architectures (with reference to pink triangle line, green left-pointing-triangle line, blue pentagram line, and brown hexagram line). 
In addition, when compared with the LLM-based model that is first trained on the 28, 15, and 5.9~GHz datasets and subsequently fine-tuned on the 1.6~GHz dataset with limited samples, WiCo-MG-Base still achieves significantly better performance, demonstrating an NMSE reduction of 3.67~dB (with reference to purple square line and pink triangle line). 
Finally, WiCo-MG-Base shows preliminary zero-shot generation capability (with reference to blue inverted-triangle line).
These findings verify that WiCo-MG-Base possesses strong generalization and data efficiency, enabling effective adaptation to new frequency scenarios with minimal data.

\subsubsection{Performance Across Different Communication Scenarios}
\begin{figure}[t]
\centering
\includegraphics[width=3.5in]{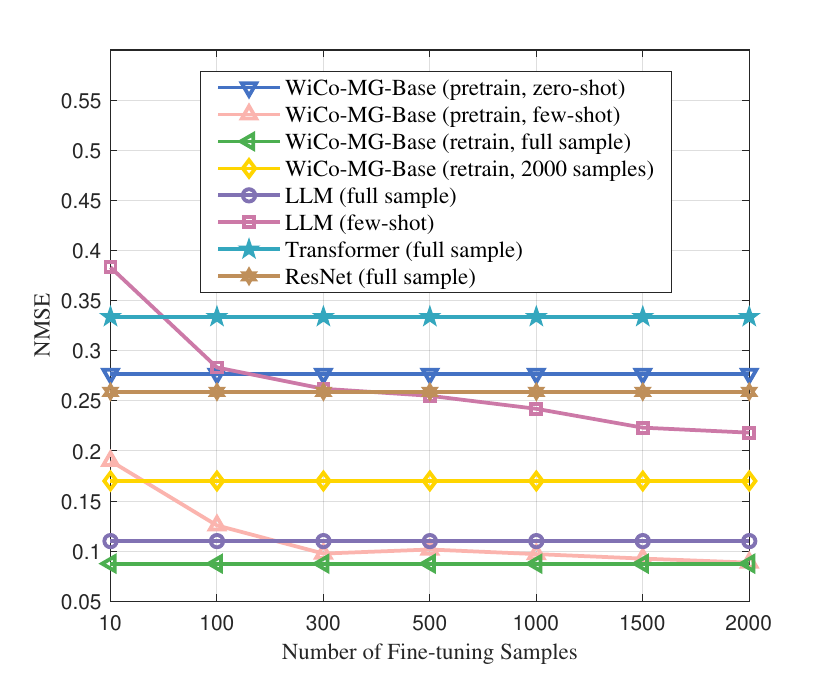}
\caption{Performance of WiCo-MG-Base and other baselines across communication scenarios.}
\label{scenarios}
\end{figure}
The experimental results across different communication scenarios are presented in Fig.~\ref{scenarios}. 
The WiCo-MG-Base is pretrained on six crossroad datasets and subsequently evaluated on the unseen wide-lane scenario to assess its cross-scenario generalization capability. 
The results demonstrate that, based on the pretraining, satisfactory performance can be achieved with only a limited number of fine-tuning samples. 
{\color{black}When approximately 1500 samples are used for fine-tuning, the performance of WiCo-MG-Base achieve a performance gap of less than 0.01 in NMSE compared with full-sample retraining (with reference to pink triangle line and green left-pointing-triangle line).}
In contrast, retraining the model from scratch with the same 1500 samples fails to reach a comparable level, indicating the effectiveness of the pretrained feature representations (with reference to yellow diamond line and pink triangle line).

Moreover, when fine-tuned with around 200 samples, the pretrained WiCo-MG-Base surpasses the full-sample trained LLM-based model on the wide-lane dataset (with reference to pink triangle line and purple circle line). 
Meanwhile, the performance of WiCo-MG-Base remains superior to that of conventional deep learning baselines, including Transformer and ResNet models trained with full samples of the wide-lane dataset (with reference to pink triangle line, green left-pointing-triangle line, blue pentagram line, and brown hexagram line). 
In addition, the LLM-based method that is first trained on the crossroad datasets and subsequently fine-tuned on the wide-lane with a few samples still exhibits a clear performance gap compared with WiCo-MG-Base. 
WiCo-MG-Base achieves an NMSE reduction of 3.75~dB (with reference to purple square line and pink triangle line).
Finally, WiCo-MG-Base exhibits preliminary zero-shot capability (with reference to blue inverted-triangle line).
These results verify that WiCo-MG-Base possesses strong generalization and adaptation capability, enabling efficient transfer from multi-scenario pretraining to new communication environments with minimal fine-tuning.

\subsubsection{Performance Across UAV Flight Altitudes}
\begin{figure}[t]
\centering
\includegraphics[width=3.5in]{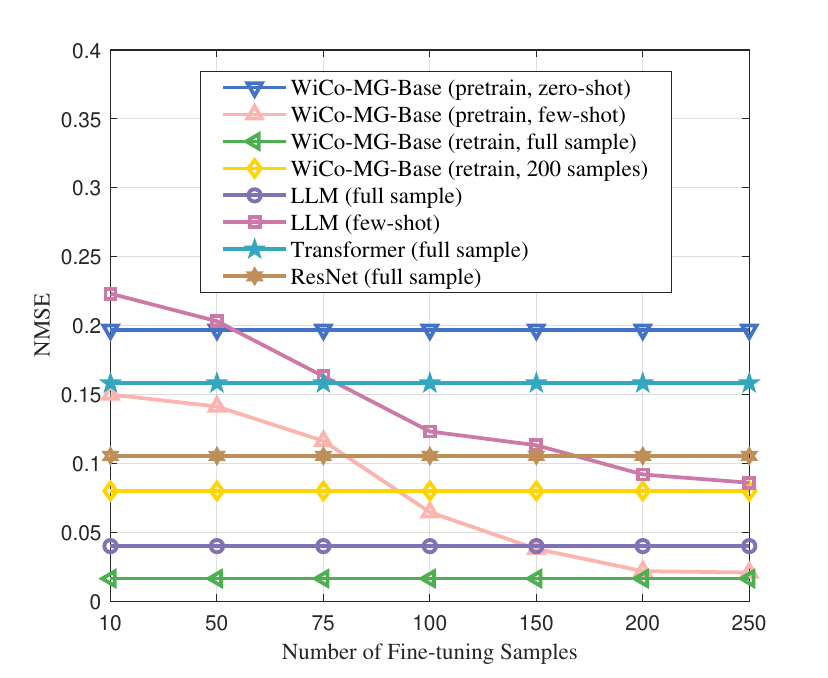}
\caption{Performance of WiCo-MG-Base and other baselines across UAV flight altitudes.}
\label{height}
\end{figure}
The cross-altitude generalization capability of WiCo-MG-Base is evaluated by pretraining the model on datasets collected at 70~m and 80~m UAV altitudes in the crossroad, followed by adaptation to the unseen 50~m dataset. 
{\color{black}The experimental results are presented in Fig.~\ref{height}. The pretrained WiCo-MG-Base demonstrates strong transferability across altitude variations, achieving performance gap of less than 0.01 in NMSE with about 200 fine-tuning samples comparable to full-sample retraining (with reference to pink triangle line and green left-pointing-triangle line).}
In comparison, retraining from scratch with the same number of samples yields inferior results, which highlights the advantage of pretraining in providing transferable representations across different altitudes (with reference to yellow diamond line and pink triangle line).

Moreover, when fine-tuned with approximately 150 samples, the pretrained WiCo-MG-Base surpasses the full-sample trained LLM-based model on the 50~m dataset (with reference to pink triangle line and purple circle line). 
{\color{black}Few-shot performance of WiCo-MG-Base with more than 100 samples and full sample performance of WiCo-MG-Base also outperform conventional deep learning baselines such as Transformer and ResNet trained with full samples (with reference to pink triangle line, green left-pointing-triangle line, blue pentagram line, and brown hexagram line).}
Meanwhile, when the LLM-based model is sequentially trained on the 70~m and 80~m datasets and subsequently adapted to the 50~m scenario with limited samples, it still exhibits a clear performance gap compared with WiCo-MG-Base (with reference to purple square line and pink triangle line).
WiCo-MG-Base achieves an NMSE reduction of 2.59~dB.
Finally, WiCo-MG-Base demonstrates preliminary zero-shot capability (with reference to blue inverted-triangle line).
These findings confirm that WiCo-MG-Base effectively captures altitude-invariant propagation characteristics through multi-altitude pretraining, enabling robust adaptation to new flight conditions with minimal fine-tuning.

\subsection{Ablation Study}
\begin{table*}[t]
\caption{Results of Ablation Experiments}
\centering
\renewcommand\arraystretch{1}
\label{Ablation}
\begin{tabular}{c|cccccc}
\toprule
                        & \textbf{D1-D12} & \textbf{D1-D6} & \textbf{D7-D12} & \textbf{D2,D4,D5,D6,D8,D10,D11,D12} & \textbf{D1,D2,D3,D7,D8,D9} & \textbf{Average} \\ \midrule
WiCo-MG-Base    &  \textbf{0.0335}      &  \textbf{0.0105}     & \textbf{0.0528} &       \textbf{0.0290}       &    \textbf{0.0425}     & \textbf{0.0340}\\ \midrule
w/o SigLIP-2      &  0.0390      &  0.0110     & 0.0542 &   0.0308      &   0.0435           & 0.0357 \\ \midrule
w/o routed expert     &  0.0842      &  0.0221     & 0.0583&  0.0320             &    0.0433    & 0.0480\\ \midrule
w/o shared expert    &  0.0648      &  0.0211     & 0.0540 &  0.0307           &     0.0441 & 0.0429\\ \midrule
w/o frequency embedding &   0.1403     &   0.0216    &  0.0959 &  0.1036               &   0.1349    &  0.0992 \\ \bottomrule
\end{tabular}
\end{table*}
To further investigate the contribution of each component in the proposed WiCo-MG-Base model, a series of ablation experiments is conducted, and the NMSE results are summarized in Table~\ref{Ablation}. 
Four configurations are examined to evaluate the influence of different modules on model performance. 
First, the SigLIP-2 module is removed to assess the effect of the continuous SoM feature space of RGB images. 
Second, the routed experts within each Transformer block are removed while retaining the frequency embedding, in order to examine the role of the routing mechanism in adaptive feature extraction. 
Third, the shared experts within each Transformer block are removed to verify their contribution to global representation learning and knowledge sharing across different scenario conditions. 
Finally, the frequency embedding in the gating network is eliminated to evaluate its impact on frequency-aware generation and overall generalization ability.

The experimental results in Table~\ref{Ablation} show that the removal of any of these components leads to a noticeable performance degradation, confirming that each module plays a crucial and complementary role in enhancing the cross-modal representation learning capability of WiCo-MG-Base.

\subsection{Scaling Analysis}
\begin{table*}[t]
\caption{Results of Scaling Analysis}
\centering
\renewcommand\arraystretch{1.5}
\setlength{\tabcolsep}{14pt}
\label{Scaling}
\begin{tabular}{ccccc}
\hline
\multirow{2}{*}{Pre-training Dadaset} & \multirow{2}{*}{Model} & \multicolumn{3}{c}{Test Dataset} \\ \cline{3-5} 
                                      &                        &   D1,D4,D9,D10      &     D2    & D8    \\ \hline
                                      & WiCo-MG-Small             &  0.0400         &    0.0210       &   0.0516       \\ \cline{2-5} 
D1-D12                                & WiCo-MG-Base              &  0.0314         &      0.0110     &    0.0476      \\ \cline{2-5} 
                                      & WiCo-MG-Large             & 0.0224  &  0.0087 &   0.0271       \\ \hline
D1,D4,D5,D6,D9,D10,D11,D12         & WiCo-MG-Base              &  0.0328         &  0.1328&      0.0708    \\ \hline
D1,D4,D9,D10               & WiCo-MG-Base              &  0.0380         &  0.1409   &   0.0762  \\ \hline
\end{tabular}
\end{table*}

Scaling analysis is essential for evaluating the impact of model capacity and pre-training dataset scale on the performance of WiCo-MG. 
As shown in Table~\ref{Scaling}, different model sizes and pre-training dataset combinations are examined, and the NMSE results are given. 
Three model variants, including WiCo-MG-Small, WiCo-MG-Base, and WiCo-MG-Large, are pretrained on datasets ranging from a subset to the complete D1-D12 datasets. 
The models are then evaluated on multiple test datasets to investigate the effects of scaling.

It can be observed that, under a fixed pre-training dataset scale, enlarging the model size from WiCo-MG-Small to WiCo-MG-Large consistently improves performance across all test datasets. 
This improvement is attributed to the enhanced representation capacity of larger models, which allows for a more comprehensive understanding of the propagation environment. 

Moreover, expanding the pre-training dataset scale substantially enhances both in-domain and cross-domain performance. 
When WiCo-MG-Base is pretrained on a larger dataset, it achieves significantly lower NMSE values on unseen datasets such as D2 and D8 compared with pretraining on smaller subsets. This result demonstrates that increasing the diversity of the pre-training data improves the model's generalization to new communication conditions.

Overall, the results confirm that the performance of WiCo-MG benefits jointly from increasing both model size and pre-training dataset scale. These findings indicate that WiCo-MG is a scalable foundation model for multipath generation, with further potential for improvement as computational resources and data availability continue to grow.

\subsection{Extensibility Analysis}
To evaluate the extensibility of WiCo-MG, two types of experiments are designed. 
The first focuses on intrinsic extensibility by training the model on multiple dominant paths simultaneously. 
The second examines post-training extensibility, where a new multipath parameter (e.g., AoA azimuth and elevation) is added to test the model's ability to adapt to new multipath parameters.
\subsubsection{Performance on the Top-N Dominant Multipath Generation}
\begin{figure}[t]
\centering
\includegraphics[width=3.5in]{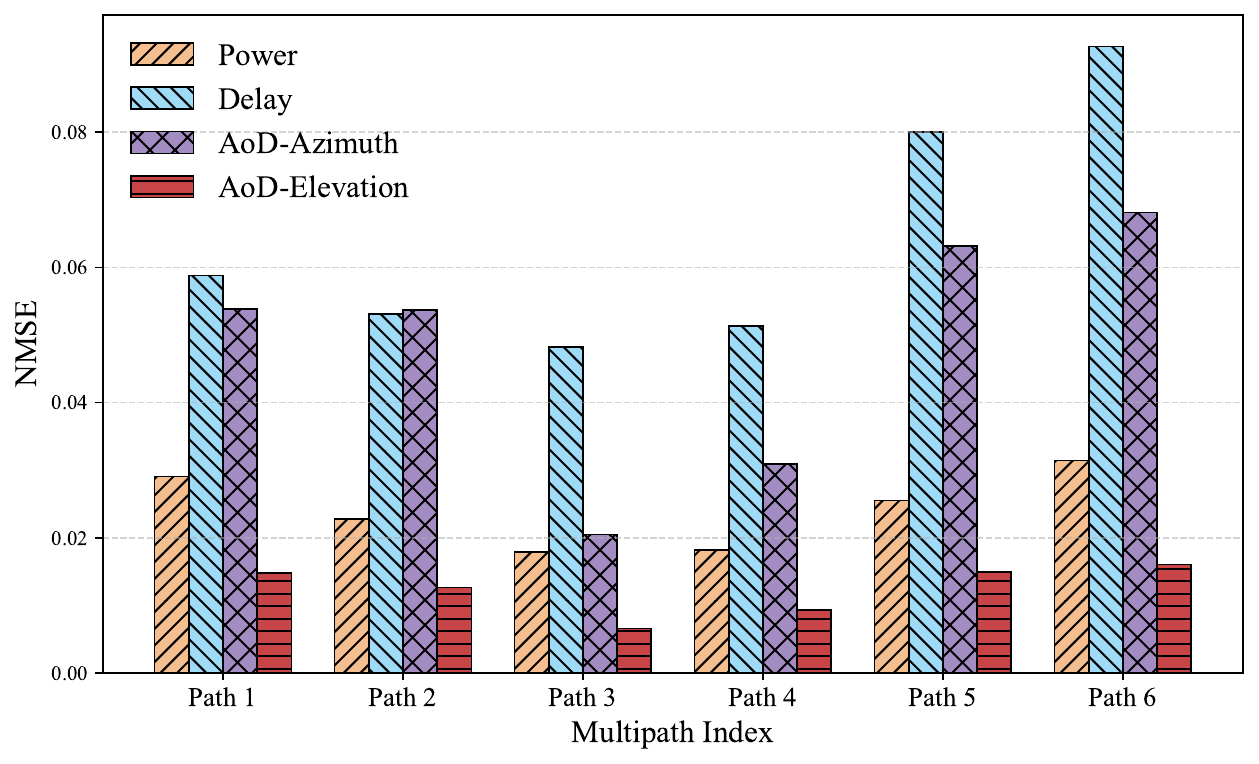}
\caption{Performance of WiCo-MG-Base on the top-N dominant multipath generation.}
\label{multiple_path}
\end{figure}

To evaluate the extensibility of the proposed model, an experiment is conducted on generating the parameters of the first several dominant paths. 
The dataset used for this study corresponds to the 28~GHz crossroad scenario at a UAV height of 70~m, which includes the power, delay, and AoD (azimuth and elevation) for the first six dominant paths. 
Specifically, the path index information is first processed through an embedding layer and a linear projection to match the dimensionality of the image tokens. 
This embedded representation is then added to all image features to inject the path-order information, while the subsequent Transformer blocks remain unchanged. Experimental results in Fig.~\ref{multiple_path} demonstrate that the average NMSE for each path's generated parameters remains below 0.06, while the overall average NMSE across all paths is below 0.04. 
These results are comparable to those obtained when the model generates only the first dominant path, demonstrating the strong extensibility of WiCo-MG in top-N dominant multipath generation.

\subsubsection{Performance with Fine-Tuning for New Multipath Parameters}\label{NewMultipath}
\begin{figure}[t]
\centering
\includegraphics[width=3.5in]{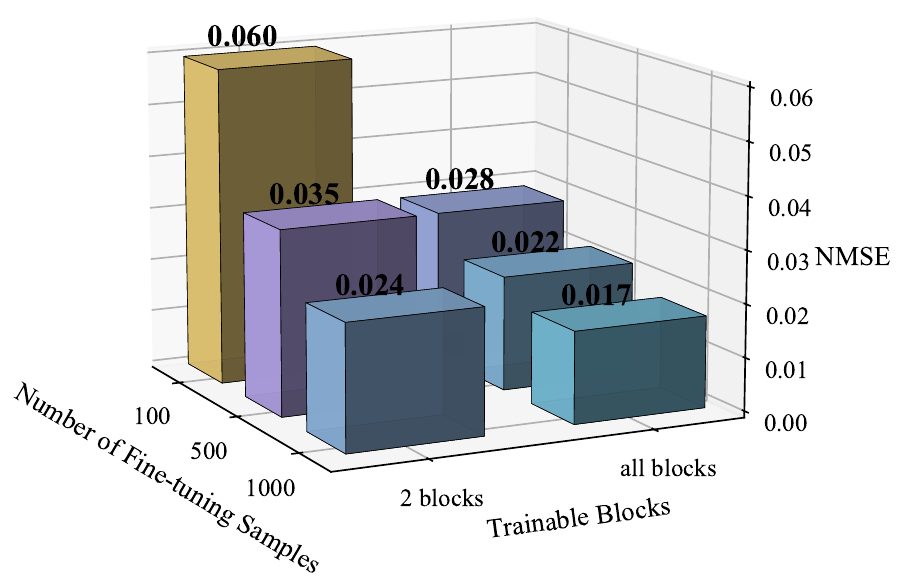}
\caption{Performance of WiCo-MG-Base on fine-tuning for new multipath parameters.}
\label{new_parameter}
\end{figure}

After training the WiCo-MG-Base, if a new multipath parameter, such as the azimuth and elevation angles of the AoA, needs to be generated, the model can be efficiently adapted through a lightweight fine-tuning process. 
Specifically, a new ViT-VQGAN is first trained for the additional parameter, and the transformer blocks that contain task-wise MoE layers are augmented with a new gating network, while the preceding token-wise MoE Transformer blocks remain unchanged. 
In the subsequent adaptation stage, one can either fine-tune the entire transformer or only the transformer blocks that include task-wise MoE layers. 
We evaluate the NMSE of new multipath parameter generation by fine-tuning a model that is originally trained on power, delay, and the azimuth and elevation angles of the AoD, to generate the azimuth and elevation angles of the AoA. 
The evaluation on AoA generation is conducted under different numbers of fine-tuning samples and varying amounts of trainable parameters to examine the adaptation efficiency of the model.
As shown in Fig.~\ref{new_parameter}, the results indicate that fine-tuning only the last task-wise MoE Transformer blocks can achieve comparable performance to full-model training once the number of fine-tuning samples reaches a certain level, while updating only about 20\% of the total parameters.

\subsection{Network Storage and Inference Cost}
The parameters of models and training and inference time cost by models are closely related to the storage and computational overhead.
The number of parameters and training and inference time of WiCo-MG-Base and the baselines are shown in Table \ref{cost}, where training and inference samples are taken from D1.
\begin{table*}[t]
\caption{Network Parameters and Inference Time Per Snapshot}
\centering
\renewcommand\arraystretch{1}
\setlength{\tabcolsep}{14pt}
\label{cost}
\begin{tabular}{c|cccc}
\toprule
                   & WiCo-MG-Base & LLM-based model & Transformer & ResNet \\ \midrule
Parameters      & 66.64M/137.90M & 129.04M&  9.81M &  13.67M \\ \midrule
Inference time &  41 ms &21 ms &    12 ms   &  16 ms  \\ \bottomrule
\end{tabular}
\end{table*}
During inference, only the RGB encoder and the multipath decoder of the ViT-VQGAN are activated, while the S-R MoE Transformer requires only a subset of experts to operate. 
This selective activation significantly reduces the total number of active parameters.
Owing to the parallelizable expert computations and independent ViT-VQGAN decoding processes, the overall inference time of WiCo-MG-Base remains within the same order of magnitude as the baselines.


\section{Conclusions}
Based on a new synthetic U2G sensing-communication dataset, WiCo-MG, a wireless channel foundation model for multipath generation via SoM, has been proposed. 
By integrating multi-modal representations of sensing and channel information, WiCo-MG has established a unified framework for cross-modal alignment and adaptive multipath generation. 
A two-stage training framework has been developed to address the challenges of representation alignment and generalization. 
The first stage has embedded sensing images and multipath parameters into coupled SoM feature spaces for modality alignment, while the second stage has employed a token-wise and task-wise S-R MoE Transformer with frequency-aware expert routing to learn flexible mappings from sensing to multipath parameters.
Experimental evaluations have demonstrated that WiCo-MG has achieved state-of-the-art in-distribution generation performance and superior out-of-distribution generalization, with an NMSE reduction of more than 2.59~dB over the baselines. 
Moreover, the model has exhibited strong scalability with increasing model and dataset sizes, and significant extensibility to new multipath parameters and tasks. 
These results have verified the effectiveness and generalization capability of WiCo-MG, highlighting its potential as a scalable foundation model for future 6G AI-native communication systems.

\section*{Funding}
This work was supported in part by the National Natural Science Foundation of China under Grant 62371273, Grant 62125101, Grant 62341101, and Grant 62571007; 
in part by the New Cornerstone Science Foundation through the XPLORER PRIZE; 
in part by the Young Elite Scientists Sponsorship Program by CAST under Grant YESS20230372.

\section*{Author Contributions}
Z. Han, L. Bai, and X. Cheng conceived the experiments, Z. Han conducted the experiments, and Z. Han, L. Bai, and X. Cai validated the results. All authors analyzed the results and reviewed the manuscript.

\section*{Competing Interests}
The authors declare no competing interests.

\end{document}